\documentclass[12pt]{article}
\usepackage{geometry}
\usepackage{graphicx}	
\usepackage{rotating}
\usepackage{amsmath}	
\usepackage{amssymb}	
\usepackage{times}
\usepackage{hyperref}
\usepackage{academicons}
\usepackage[dvipsnames]{xcolor}
\usepackage[superscript,biblabel]{cite}
\usepackage{url}
\usepackage{etoolbox}
\usepackage{caption}

\title{Identification of plasma modes in Galactic turbulence with synchrotron polarization}
\author{Heshou Zhang$^{1,2}$, 
  Alexey Chepurnov$^{1}$,
  Huirong Yan$^{1,2 \ast}$, 
  Kirit Makwana$^{1}$, \\
  Reinaldo Santos-Lima$^{1,2}$ and
  Sarah Appleby$^{3}$}

\begin{document}
\label{firstpage}
\maketitle

  \begingroup
  \renewcommand\thefootnote{}\footnote{$^{1}$Deutsches Elektronen-Synchrotron DESY, Platanenallee 6, D-15738 Zeuthen, Germany.
$^{2}$Institut f$\ddot{u}$r Physik und Astronomie, Universit$\ddot{a}$t Potsdam, Haus 28, Karl-Liebknecht-Str. 24/25, D-14476 Potsdam, Germany.
$^{3}$Institute for Astronomy, University of Edinburgh, Royal Observatory, Blackford Hill, Edinburgh, EH9 3HJ, UK. $\ast$Email: huirong.yan@desy.de.}%
  \addtocounter{footnote}{-1}%
  \endgroup

{\bf
Magneto-hydrodynamic (MHD) turbulence is ubiquitous and a fundamental ingredient underlying many astrophysical phenomena. The multiphase nature of interstellar medium and diversity of driving mechanisms give rise to spatial variation of turbulence properties, particularly their plasma properties. There has been no observational diagnosis of the plasma modes so far beyond the solar system.
Here we report the identification of different plasma modes in various Galactic environments, including active star forming zones and supernova remnants, based on our synchrotron polarization analysis. The observed high degree of consistency between the $\gamma-$ray excess in Cygnus cocoon and the location of magnetosonic modes provides strong observational evidence for the long-advocated theory that magnetosonic modes dominate the cosmic ray (CR) scattering and acceleration. Our results open up a new avenue for the study of interstellar turbulence and demonstrate the indispensability to account for their plasma property in all the relevant processes including CR transport and star formation.
}

Most of the electromagnetic perturbations in astrophysics are in the form of plasma turbulence because of the vast size span of astrophysical systems and the resulting huge Reynolds number. The interstellar medium (ISM) is turbulent on scales ranging from AUs to hundreds of parsecs\cite{Armstrong95, Elmegreen2004}{}, with an embedded magnetic field that influences almost all of its properties, including thermal conductivity, wave propagation and CR transport as well as star formation. The rough equipartition of magnetic and thermal energy in the interstellar medium indicates the magneto-hydrodynamic (MHD) nature of the interstellar turbulence. The apparent paradox is: on one hand, it is recognized that interstellar magnetic field is turbulent; on the other hand, the turbulence is described as hydrodynamic turbulence with isotropic Kolmogorov scaling. Most observational studies of turbulence focus on the one dimensional spectrum in the inertial range. This is inadequate, however, in the case of MHD turbulence, which has three-dimensional (3D) structures and anisotropies.
Unlike hydrodynamic turbulence, MHD turbulence can be decomposed into three plasma modes with different dispersion relations and characteristics. Particularly, the Alfv\'en and slow modes have scale-dependent anisotropy\cite{GS95, LG01}{}. In comparison, fast modes are much more isotropic \cite{CL03, MY19}{}. As a result, the scattering of charged particles by Alfvenic turbulence is negligible and magnetosonic (MS) modes are found dominant for processes such as CR transport and acceleration \cite{YL02, Lynn14}{}.

There are a variety of the drivers for turbulence in ISM ranging from supernovae explosions \cite{MO77}{}, accretion flows\cite{KH10_accretion}{}, magneto-rotational instability in the galactic disk \cite{Balbus99}{}, thermal instability \cite{KN02_thermal}{}, to collimated outflows \cite{NL07_outflow}{}, etc. The diversity of driving mechanisms and multiphase nature of ISM naturally give rise to spatial variations of turbulence properties, in particular, the relative proportion of the three modes. Nevertheless, the employed model of interstellar turbulence is often oversimplified, assumed to be only Alfv\'enic or even hydrodynamic due to a lack of observational evidence.  One argument has been that MS modes are subjected to severe damping. However, quantitative studies have demonstrated that the impact of damping is limited. The cascade of fast modes, for instance, can survive below sub-parsec scales in fully ionized plasma\cite{YL08}{}.

It is a great challenge, however, to remotely diagnose the MS modes in turbulence. The only exception is the solar wind, where slow modes have been directly detected \cite{slowmodes_shi15}{}. Here we report the employment of our novel method, the signature from polarization analysis (SPA), on unveiling the plasma modes in interstellar turbulence. The diffuse synchrotron radiation is generated when relativistic electrons travel in the magnetic field. The signatures of the plasma properties in interstellar magnetic turbulence are thus coded in the synchrotron radiation, especially its polarization signals. In particular, we utilize the fact that MHD turbulence is statistically axis-symmetrical with respect to the direction of local mean magnetic field. Through both analytical study and synthetic observations on various turbulence data generated from MHD simulations, the SPA method has established the link between the synchrotron polarization properties and the underlying turbulence statistics associated with different plasma modes. The SPA method is then applied to the synchrotron polarization data from two types of galactic medium with prominent synchrotron radiation. The identified modes in these regions are compared to multi-wavelength maps to illustrate the underlying physical processes generating the corresponding turbulence modes and their connections with related processes, such as CR transport.

\section*{Results}

The statistical properties of the Galactic magnetic turbulence are imprinted in the emissivity of the polarized synchrotron radiation ($\varepsilon_{x}$), which can be expressed by the Stokes parameters $\varepsilon_{x}\equiv(I+Q)/2$. Here, $x$ refers to the $x-$axis of the Stokes frame and $I\propto(B^2_{x_s}+B^2_{y_s})$, $Q\propto(B^2_{x_s}-B^2_{y_s})$ (see Figure 1a for definitions of angles, the electron index is assumed to be 3.0). Hence, the emissivity can be expressed by $\varepsilon_{x}({\bf{R}})= B_{0\bot}^2\cos^2\varphi_s + 2B_{0\bot}\cos\varphi_s B_i \hat{e}_{{x}_s{i}} + (B_i \hat{e}_{{x}_s{i}})^2$, where $\varphi_s$ is the azimuthal angle of the mean polarization direction measured from the Stokes parameter frame, $B_{0\bot}$ and $B_i$ are the local mean and fluctuating magnetic field, respectively. The angle-independent factors are ignored. In order to reveal the statistics in the magnetic fluctuations, a natural and intuitive quantity is the variance. In our method, the variance of the synchrotron emissivity is adopted as the modes classification signature:
\begin{equation} \label{eq:sxx_var}
\begin{array}{ll}
s_{xx} = {\it Var}(\varepsilon_x)
  &\equiv \left<{\varepsilon_{x}^2}\right>-\left<{\varepsilon_{x}}\right>^2 \\
  & \propto \left[\int C_{xx}({\bf{R}}) - \int C_{xx}({\bf{0}})\right] d{\bf{R}} \simeq (2B_{0\bot}\cos\varphi_s)^2 \int d{{\bf{k}}} F^2_{xx, l}({\bf{K}}),
\end{array}
\end{equation}
where $C_{xx}$ is the correlation function of the emissivity. $F^2_{xx, l}({\bf{K}})$ is the linear term of the tensor power spectrum for magnetic turbulence. When linear plasma modes dominate, the magnetic turbulence is axisymmetric and the power spectrum can be decomposed into a combination of two spectral tensors: $T_E$ (isotropic) and $T_F$ (anisotropic), both of which can be expressed in the format of ``$a\sin^2\varphi_s$+b'' (see Method for details). As a result, our classification signature comes down to a function of $\varphi_s$:
\begin{equation}
s_{xx}(\varphi_s) = (a_{xx}\sin^2\varphi_s+b_{xx})\cos^2\varphi_s, \; \varphi_s \in [0,\pi].
\label{eq:sxx}
\end{equation}
The quantities $a_{xx},{b}_{xx}$ carry information on the tensor power spectrum of the turbulence modes\cite{CL03, YL04, LP12}{}. The ratio ${r}_{xx}\equiv {a}_{xx}/{b}_{xx}$ is the classification parameter in SPA. It can be analytically expressed depending on three physical parameters (see Supplementary and Extended Data Fig. 1a$\sim$c): the angle $\theta_\lambda$ between the mean magnetic field $B_0$ and the line of sight; Alfv\'enic Mach number $M_A\equiv\delta{V}/v_A$, where $\delta{V}$ depicts the velocity fluctuations on the injection scale of turbulence and $v_A$ is the Alfv\'en speed; and plasma $\beta$ which is the ratio of thermal to magnetic pressure.
The value range of ${r}_{xx}$ varies with different plasma modes.

\paragraph*{Numerical simulations}

The classification recipe is further established from numerical analyses. 3D magnetic turbulence data cubes are generated with various $M_A$ and plasma $\beta$ from MHD simulations with different state-of-the-art MHD codes with both solenoidal (PENCIL \cite{PENCILCITE}{}, PLUTO \cite{PLUTOCITE}{}) and compressible (PLUTO) driving being implemented.
The synthetic observation is carried out by analyzing the polarized synchrotron radiation from the turbulence data cubes through randomized lines of sight with ray-tracing method (see Figure 1b, Method). The corresponding $s_{xx}(\varphi_s)$ is calculated by gradually rotating the Stokes frame on the picture plane from the mean polarization direction for a range of $180^\circ$, effectively covering $\varphi_s \in [0,\pi]$ (see Figure 1a right). We only accept the $s_{xx}$ where the linear component is dominant, i.e., $\xi_{xx}\equiv Amp_{linear}/Amp_{total}>0.5$. Additionally, given synchrotron polarization direction is perpendicular to the magnetic field projection on the plane of sky, the local mean magnetic field is at $\varphi_s=90^\circ$. Only the signatures axisymmetric to $90^\circ$ are accepted, therefore featuring the linear plasma modes. The resulting $r_{xx}$s produced from each case are evaluated statistically with the probability distribution analysis (PDA). There are three types of distinguishable signatures based on the range of $r_{xx}$ and their linear component amplitude: (1) ``Ambiguous'' $-1\leq r_{xx}\leq -1/3$; (2) ``MS'' $-1/3\leq r_{xx}\leq 0$; and (3) ``Alfv\'enic'' $r_{xx}>0 + \xi_{xx}<0.95$. The numerical results of PDA are presented in the following paragraph (see Methods for details).

The first tests were conducted with  the decomposed modes from those 3D MHD turbulence data cubes. The Alfv\'en and MS modes are found to be well distinguishable by the dominance of their corresponding signatures (see Extended Data Fig. 1d$\sim$f). Note that the magnetic data cubes from decomposed modes cannot represent real ISM magnetic turbulence.
We study then the whole turbulence data cubes directly from MHD simulations, whose energy, based on different driving mechanism (solenoidal or compressible forcing), differs in the composition of Alf\'en modes and MS modes\cite{MY19}{}. The synchrotron polarization signals observed from real galactic environments are subjected to Faraday rotation (FR) while propagating through the magnetized ISM. The influences from two types of FR are accounted in our simulations: the FR within the emitting layer (Figure 2a,b) and the FR in a foreground screen (Figure 2c,d).  The influence from the foreground FR is found to be marginal. The dominant signature reveals the prominent modes if the rotation angle $\theta_{FR}\lesssim 45^\circ$ within the emitting layer. Additionally, the numerical tests were also performed for $\gamma\sim2.5$. The resulting signatures with non-rotation, $\theta_{FR}=45^\circ$ FR within emitting layer, as well as $\theta_{FR}=45^\circ$ FR through foreground screen are shown in Figure 2e. We conclude that the dominant signature unambiguously reveals the corresponding modes.
Finally, our classification recipe is established as summarized in Figure ~\ref{obs_illus}a .

\paragraph*{Observational analysis}

We use the synchrotron polarization data from ``Urumqi $6~{\rm cm}$'' polarization survey with the beam size $9.\mkern-4mu^\prime5$ \cite{Xiao2011}{}. The maps are denoted by their Galactic coordinates: ($l; b$), where $l$ is the longitude and $b$ is the latitude. Two regions on the sky are analyzed: Cygnus X and the region near Rosette Nebula. All the relevant scales are listed in the corresponding figures.

In the numerical simulations, the white noise is introduced into the our simulated synchrotron polarization signal with the noise level similar to the sensitivity of the Urumqi survey and then beam smoothing is performed on the resulting polarization map by convolving with a Gaussian beam of the size of the telescope resolution (see Methods for details). As demonstrated in Figure 2f, the correspondence between the dominant signature and the corresponding modes are unaffected under current telescope sensitivity and resolution, verifying the classification recipe as illustrated in Figure 3a.

When implementing SPA on the synchrotron polarization map from real observation, we adopt the acceptance criteria with the same filters as in numerical simulations for linear dominance $\xi_{xx}>50\%$ and axisymmetry (see Methods for detail). $s_{xx}(\varphi_s)$ is calculated for a ``spot'', whose size is $\lesssim$ the expected injection scale of interstellar turbulence. In this study, the turbulence energy injection scale are of the size {L} $\lesssim 100pc$. In each spot, Fourier filtering is performed on the observed polarization map. The smallest scale is limited by the telescope beam size, which is $\sim 4pc$.
The large scale filtering scans scales from the spot size to gradually decreasing scales, in order to search for the dominant signature for different plasma modes. The purpose of this procedure is to ensure adequate statistics for the eddies. Here, the prominent signatures is found on scales of a few tens parsecs for both regions under study. Note that the results are dominated by the properties of the large scale eddy.

Furthermore, spot-by-spot analysis is conducted across the map. PDA is then conducted on a cluster of spots, all of which cover the same eddy at the center. The cluster core, the overlapping zone of all the spots in the cluster, with a radius of half the spot size, is the unit to denote the modes property.
The classification threshold is set as $75\%$ dominance of one signature over all the accepted spots (see more details in Methods section and Figure~\ref{obs_illus}b).

For each cluster in the sky map in our analysis, the statistics from more than 300 spots covering it are analyzed. The statistics for the observational analysis, i.e., the percentage of MS and Alfv\'enic signatures for different clusters, are plotted in Extended Data Fig. 2. The regions identified with dominant signatures are then compared with the synchrotron intensity maps and the FR measured through the whole Galaxy \cite{Opper2015}{}. In addition, the FR of Cygnus X region is estimate by considering a typical electron density of warm ionized medium of $1 cm^{-3}$ and magnetic field of $5\mu{G}$ with a depth of $200 pc$, which yields $\theta_{FR}=RM\lambda^2\simeq 8^\circ$. The FR within the emitting layer has therefore little influence on the SPA for the Cygnus X region.

The results of the modes identification with SPA are displayed in Figure~\ref{Cyg} for Cygnus X region, a complex of giant molecular clouds hosting massive star-forming activities with rich collection of young massive stars and supernovae \cite{Beerer10,Herschell16,Wright12}{}. The signatures superimposed on the synchrotron intensity indicate that the plasma modes are from the region rather than the foreground, since no signature is detected in low intensity zone (see Figure~\ref{Cyg}a). Moreover, the signatures are plotted over the FR map in Figure~\ref{Cyg}b. The signatures which can be influenced by FR, i.e. with rotation measure $\theta_{FR}> 45^\circ$, are all removed.

The identified signatures are also overlaid on the Extinction map (see Figure~\ref{Cyg}c). The overall detected plasma modes overlap to a large extent with the important active star forming regions, Cygnus X South\cite{Schneider2011,Rygl12,Maia16}{}. The middle 2-degree zone exhibits substantial amount of MS modes. Alfv\'en modes are also discovered in the north and south regions.

Moreover, this region has also a diffuse Fermi superbubble of $\gamma-$ray excess above 3GeV (Figure~\ref{Cyg}d)\cite{FermiLAT:2011}{}, which can be explained neither by neighboring pulsar wind nebulae nor by density enhancement as indicated by CO map (see Figure~\ref{Cyg}e). In addition, intense CR emission in Cygnus cocoon is also detected by HAWC\cite{HAWCCYG}{}. Figure~\ref{Cyg}f clearly demonstrates that the Cocoon is correlated with the identified magnetosonic modes to a high degree of consistency. Interstellar turbulence has a huge span ranging from $\sim 100$pcs to $\sim 10^9$cms as observed \cite{Armstrong95}{}. The Alfv\'enic turbulence and MS modes become decoupled on scales smaller than the injection scale and form separate cascade \cite{CL03}{}. Consequently, the magnetosonic perturbations discovered on the scales of a few tens parsecs indicate that the percentage of magnetosonic modes can be much higher than other regions on all smaller scales down to dissipation. Earlier studies demonstrate that the MS modes play a dominant role in CR scatterings \cite{YL02, YL04}{}, thus providing a stronger confinement for CRs. Therefore, our observation of plasma modes unveils the origin of the CR concentration in the Cygnus cocoon, which is completely in line with the theoretical predictions. The plasma modes information is evidently indispensable in the studies of CR propagation and acceleration.

We also performed SPA in the vicinity of Rosette Nebula (see Figure~\ref{SNR}). As demonstrated in Figure~\ref{SNR}b, the Galactic FR is low in the outer Galactic disk. Previous study has shown that the Rosette Nebula and the supernova remnants (SNR) ${\rm G205.5+1.5}$ are located at the same distance, interacting with each other \cite{Odegard1986}{}. The Alfv\'enic signatures seen at the western and northern edges of the Rosette Nebula point to the Alfv\'en modes since the FR angle here is $\lesssim 30^\circ$. Furthermore, MS modes is only observed in the center of the SNR. The MS and Alfv\'en signature dominance imply their corresponding forcing mechanisms. It is plausible that the turbulence within the MS cluster is driven compressively by the supernova shock, leading to substantial amount of compressible MS modes. On the other hand, the turbulence in the edges of the molecular cloud is probably driven by the shearing (solenoidal) motion, resulting in the dominance of incompressible Alfv\'en modes as observed. In addition, the prominent isotropic clusters emerged in between the SNR and molecular cloud at ($206, -0.5$), indicating the existence of a super-Alfv\'enic turbulence.

The application of SPA on the synchrotron polarization data from the Galactic medium has for the first time revealed that interstellar turbulence is magnetohydrodynamic with different plasma modes composition, pinpointing the necessity to account for the plasma properties of MHD turbulence, which is neither hydrodynamic nor purely Alfv\'enic, but depends on local physical conditions, particularly the driving process. This study opens up a new avenue of connecting plasma physics with macroscopic astrophysical phenomena. Moreover, the observation of plasma modes composition in the Galaxy has far reaching consequences on not only CRs and star formation, but also the fundamental understanding of the driving mechanism of turbulence. Further investigation of the statistical properties of the signatures in SPA will allow robust estimates of magnetic field direction and strength, Alfv\'enic and sonic Mach numbers and the energy injection scale of turbulence. A highly promising research field is foreseen to unroll with ample results anticipated from the advanced analysis of high resolution synchrotron polarization data and multiple wavelength comparison, that will shed light on the role of turbulence in various physical processes.

\section*{Acknowledgments}
We thank the helpful communications on various topics discussed
in this paper with the following colleagues: F. Boulanger, M. Gangi, S. Gao, A. Lazarian, H. B. Liu, R. Liu, J. Liu, M. Pohl, I. Sushch, A. Taylor, J. Volmer, M. Vorster, X. Wu, Q. Zhu. S.A. acknowledges the support from the DESY summer student program.

\paragraph*{Author Contributions}
H.Y. oversaw the project. A.C., H.Y., and H.Z. contributed to the theoretical analysis of SPA. K.M. and R.S.L. prepared the turbulence data. H.Z. carried out the numerical simulation, performed the probability distribution analysis on the simulation results and led the establishment of the SPA recipe. A.C., S.A. and H.Z. analyzed the data from real observations. H.Z. and H.Y. interpreted the observational results and led the manuscript preparation. H.Y. and A.C. designed the project. All authors discussed the project and commented on the manuscript.

\paragraph*{Competing interests}
The authors declare no competing financial interests.

\paragraph*{Additional Information} {\bf Correspondence and requests for materials} should be addressed to H.Y.

\newpage

\begin{figure}
\begin{center}
\includegraphics[width=0.75\textwidth]{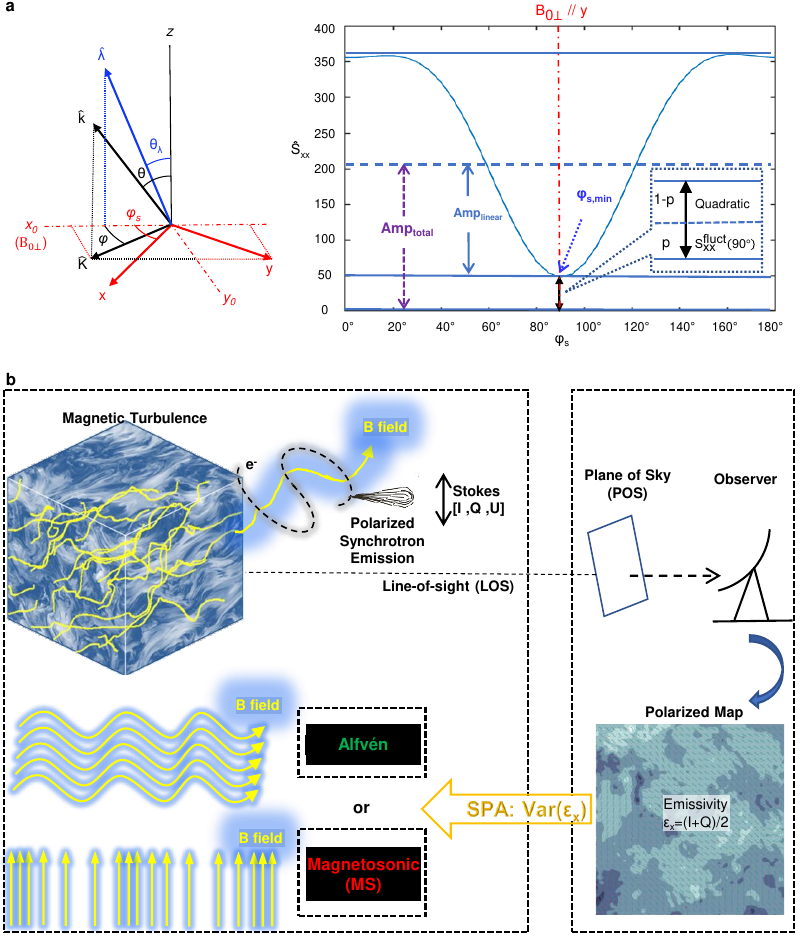}
\end{center}
\caption{{\bf $\mid$ Illustration for SPA recipe.} (a) {\it left:} Vectors and angles involved in the calculation of the spectral tensor. The axis $Oz$ points to the observer. $xOy$ is the plane of sky (POS). ${\bf \hat{\lambda}}$ is the direction of the mean magnetic field, of which the POS component defines the $y_0-$axis. The angle between the mean field direction and LOS is $\theta_\lambda$. ${\bf\hat{k}}$ is the wave vector and ${\bf\hat{K}}$ is its POS component. $\varphi$ is the angle between ${\bf\hat{K}}$ and $x-$axis. The angle between the wave vector ${\bf\hat{k}}$ and LOS is $\theta$. Vectors $x$ and $y$ define the frame of the Stokes parameters. $\varphi_s$ is the positional angle of $B_{0\bot}$ measured in the Stokes parameter frame. {\it right:} The composition of the raw signature $\hat{s}_{xx}$ from direct observation. The local mean field on POS, $B_{0\bot}$, is parallel to the Stokes $y$ axis when $\varphi_s=90^\circ$. The spot is accepted under axisymmetric criteria (see Methods). The amplitudes for the linear part of signature $Amp_{linear}$ and total signature $Amp_{total}$ are marked. The non-zero value of $\hat{s}_{xx}(90^\circ)$ consists of three parts: the quadratic term, free-free emission and the symmetry axis fluctuation $s^{fluct}_{xx}$, where the percentage of fluctuation term contribution is $p$.(b) The schematics for SPA analysis on observations. Relativistic electrons produce the polarized synchrotron emission that carries the plasma statistics, yielding the synchrotron polarized map on POS. Reciprocally, the variance of synchrotron emissivity in the observed map can be used to recover the plasma modes information in the original turbulence. }
\end{figure}

\begin{figure}
\begin{center}
\includegraphics[width=1\textwidth]{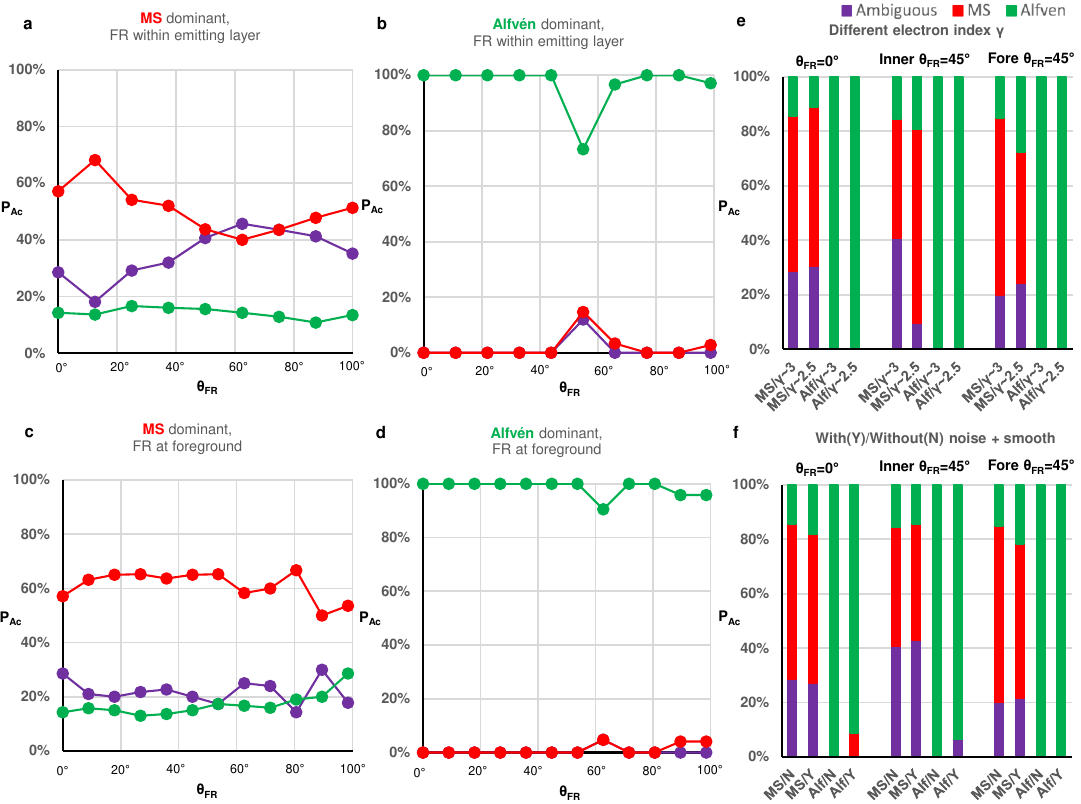}
\end{center}
\caption{{\bf $\mid$ Numerical simulations on total MHD turbulence data cubes with different composition of plasma modes.} $(a\sim b)$: tests accounting for the Faraday rotation (FR) within the emitting layer; $(c\sim d)$: tests accounting for FR as synchrotron radiation transports through a foreground screen. $\theta_{FR}$ is the rotational angle. y-axis $P_{Ac}$ is the percentage of different signatures among the accepted testing results. The color code is marked on the upper right: {\color{ForestGreen}Green}: ``Alfv\'enic''; {\color{red} Red}: ``MS''; {\color{Purple}Purple}: ``Ambiguous'' (see Methods). (e) Comparison between electron indices $\gamma\simeq 2.5, 3.0$ without and with two types of FR counted. (f) Comparison between simulations with (Y) and without (N) noise and beam smoothing, based on the telescope configuration for the Urumqi survey (see Methods for details). All numerical tests have revealed that the dominant signature unambiguously reveals the corresponding plasma modes.}
\label{num_res}
\end{figure}

\begin{figure}
\begin{center}
\includegraphics[width=1\textwidth]{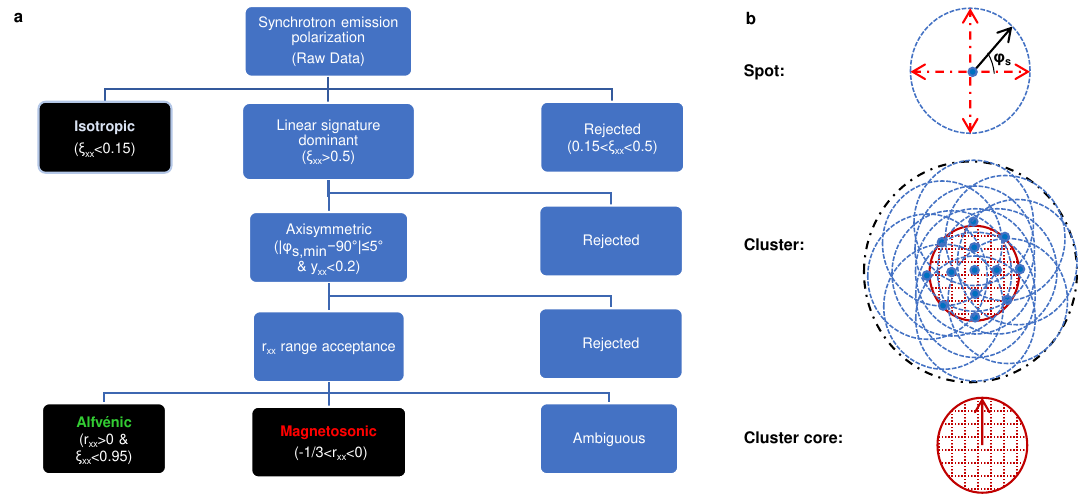}
\end{center}
\caption{{\bf $\mid$ Applying SPA to observational data} (a) Schematics of the modes classification from the synchrotron polarization map with the ``SPA'' code.(b) Cluster analysis. 1) The $s_{xx}$ is calculated within a spot with respect to its local mean magnetic field $B_{0\bot}$; 2) A cluster of spots are analyzed; 3) The centres of all the spots considered reside in the overlapping core of the cluster, which is half the spot size. The percentage of modes signature is defined by the ratio between the number of the detected signature and the total amount of accepted spots within a cluster. The resulting percentage of different modes signatures for the two regions under study is presented in Extended Data Fig. 2.}
\label{obs_illus}
\end{figure}

\begin{figure}
\begin{center}
\includegraphics[width=1\textwidth]{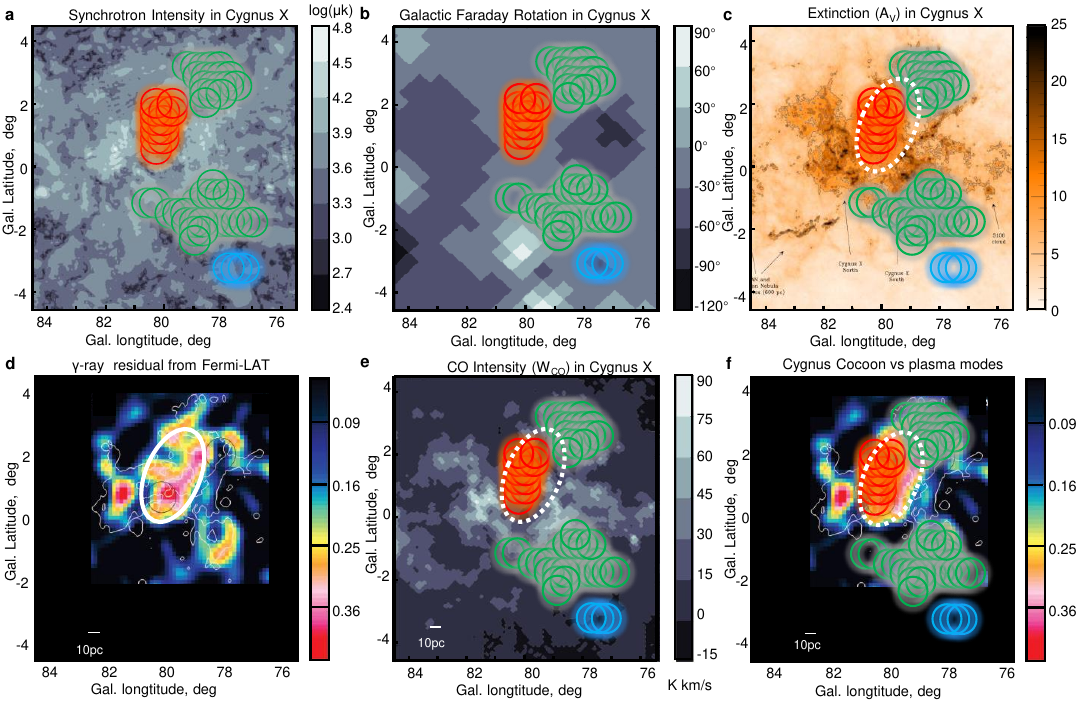}
\end{center}
\caption{{\bf $\mid$ Turbulence modes identified in Cygnus X region} The detected clusters are plotted over the {\em a}) synchrotron intensity map and {\em b}) FR map. The color code for signatures: {\color{ForestGreen}Green}: ``Alfv\'enic''; {\color{red} Red}: ``MS''; and {\color{Blue} Blue}: ``isotropic turbulence''. The distance of the object is $1.4kpc$. The spot size is $50pc$. The size of the largest eddy is $\sim 15pc$. Furthermore, turbulence modes identified in Cygnus X region are plotted over {\em c}) the Near-Infrared extinction map \cite{Schneider2011}{} where the high-mass star-formation are indicated. {\em d}) An illustration for the photon count residuals of $\gamma-$ray emission from Fermi-LAT showing the extended emission excess, the "Fermi cocoon" (in white circle) \cite{FermiLAT:2011}{}. {\em e}) Turbulence modes identified in Cygnus X region plotted over the CO intensity contour map ($W_{CO}$ in $K km s^{-1}$ unit)\cite{COCompare}. {\em f}) Comparison between Fermi-LAT $\gamma-$ray residuals and the identified turbulence modes. The region identified with MS modes overlaps in a high consistency with the "Fermi cocoon".}
\label{Cyg}
\end{figure}

\begin{figure}
\begin{center}
\includegraphics[width=1\textwidth]{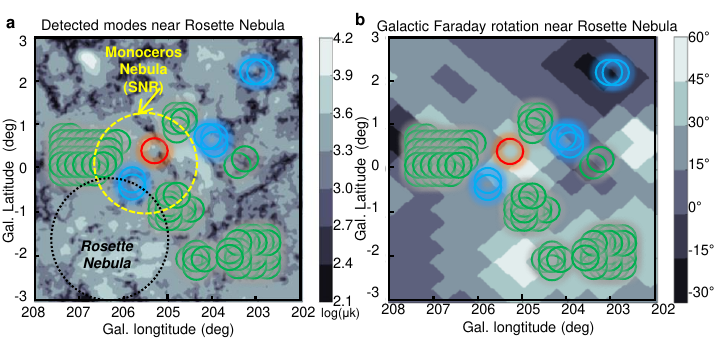}
\end{center}
\caption{{\bf $\mid$ SPA analysis in the vicinity of Rosette nebula.} The color codes for the identified signatures are the same as Fig. 4. {\em a}) The detected signatures compared with synchrotron intensity. The center of the map is SNR G205.5$+$0.5. Bottom left is Rosette Nebula. {\em b}) The Galactic FR of the area. The distance to both the nebula and the SNR is $\sim1.6kpc$. The spot size is $36pc$. The scale of the largest eddy is $\sim18pc$. Isotropic clusters are identified between the SNR and the Rosette nebula with high synchrotron intensity.
}
\label{SNR}
\end{figure}

\clearpage

\section*{Methods}
\paragraph*{Theoretical Analysis}

We first derive the relation between the variance of emissivity and correlation function, then use spectral representation to express the analytical expectation of the signature. The synchrotron emissivity related to (I+Q)/2 is $\varepsilon_{x} = B_{0\bot}^2\cos^2\varphi_s + 2B_{0\bot}\cos\varphi_s B_i \hat{e}_{{x}{i}} + (B_i \hat{e}_{{x}{i}})^2$, where $B_{0\bot}$ is the projection of the mean magnetic field in the plane of sky and $B_i$ is the turbulent magnetic field. $\varphi_s$ is the positional angle of $B_{0\bot}$ in the Stokes parameter frame (see Figure 1a).

The correlation function of the emissivity at two points separated by ${\bf{r}}$ is
\begin{equation}
C_{xx}({\bf{r}}) \equiv\left<{\varepsilon_{x}({\bf{r}})\varepsilon_{x}({\bf{0}})}\right> = \frac{\int d{\bf{p}}\int d{\bf{q}}\varepsilon_{x}({\bf{p}})\varepsilon_{x}({\bf{q}})\delta({\bf{p}}-{\bf{q}}-{\bf{r}})}{\int d{\bf{p}}\int d{\bf{q}}}\end{equation}
where ${\bf{p}}$, ${\bf{q}}$ are two random points from the space. Integrated over all distances:
\begin{equation} \label{eq:cxx_int}
\int C_{xx}({\bf{r}}) d{\bf{r}}
  = \int d{\bf{r}}\frac{\int d{\bf{p}}\int d{\bf{q}}\varepsilon_{x}({\bf{p}})\varepsilon_{x}({\bf{q}})\delta({\bf{p}}-{\bf{q}}-{\bf{r}})}{\int d{\bf{p}}\int d{\bf{q}}} = \left<{\varepsilon_{x}}\right>^2
\end{equation}
Therefore, ${\it Var}(\varepsilon_{x})= \int d{\bf{r}} \left[C_{xx}({\bf{r}})-C_{xx}({\bf{0}})\right]$.

A random phased model of the turbulent magnetic field can be described with the following spectral representation in Fourier space \cite{R90,Che98}{}:
\begin{equation} \label{eq:B}
B_i({\bf{r}}) = \int \sqrt{d{\bf k}} e^{i{\bf{k}}\cdot{\bf{r}}} F({\bf{k}}) \, T_{ij}({\bf \hat{k}}) \xi_j({\bf{k}}) .
\end{equation}
\noindent Here {\bf r} is the position vector, ${\bf\hat{k}}$ is the wave vector. The spectral tensor $T$ and complex random field $\xi$ obey the following rules: $T_{ij} = T_{il}T_{jl}$, $\left<{\xi_i({\bf{k}})\xi_j^*({\bf{k}}')}\right> = \delta_{ij} \delta_{{\bf{k}}{\bf{k}'}}$ and $\xi_i(-{\bf{k}}) = \xi_i^*({\bf{k}})$. The brackets ``$\left<,\right>$'' here designate the ensemble averaging. $F({\bf{k}})$ is the square root of the scalar factor in the tensor spectrum.

Using (\ref{eq:B}), we can derive the following expression for the emissivity correlation function:
\begin{equation} \label{eq:cxx}
\begin{array}{lll}
C_{xx}({\bf{r}})
  & \equiv & \left<{\varepsilon_{x}({\bf{r}})\varepsilon_{x}({\bf{0}})}\right> \\
  & = & \left<\left(B_{0\bot}^2\cos^2\varphi_s
    + 2B_{0\bot}\cos\varphi_s      \int \sqrt{d{\bf{k}}_{1}} e^{i{\bf{k}}_{1}{\bf{r}}} F({\bf{k}_{1}}) \, \hat{e}_{x_si_1} T_{i_1 j_1}({\bf \hat{k}}_1) \xi_{j_1}({\bf{k}}_1) \right.\right. \\
  & + &                  \int \sqrt{d{\bf{k}}_{2}} e^{i{\bf{k}}_{2}{\bf{r}}} F({{\bf{k}} _{2}}) \, \hat{e}_{x_si_2} T_{i_2 j_2}({\bf \hat{k}}_2) \xi_{j_2}({\bf{k}}_2) \\
  & \; \cdotp & \left.\! \int \sqrt{d{\bf{k}}_{3}} e^{i{\bf{k}}_{3}{\bf{r}}} F({{\bf{k}} _{3}}) \, \hat{e}_{x_si_3} T_{i_3 j_3}({\bf \hat{k}}_3) \xi_{j_3}({\bf{k}}_3) \right) \\
  & \; \cdotp & \left(B_{0\bot}^2\cos^2\varphi_s
    + 2B_{0\bot}\cos\varphi_s               \int \sqrt{d{\bf{k}}'_{1}} F({{\bf{k}}'_{1}}) \, \hat{e}_{x_si'_1} T_{i'_1 j'_1}({\bf \hat{k}}'_1) \xi_{j'_1}({\bf{k}}'_1) \right. \\
  & + &                           \int \sqrt{d{\bf{k}}{'}_{2}} F({{\bf{k}}{'}_{2}}) \, \hat{e}_{x_si'_2} T_{i'_2 j'_2}({\bf \hat{k}}'_2) \xi_{j'_2}({\bf{k}}'_2) \\
  & \; \cdotp & \left.\left. \!\! \int \sqrt{d{\bf{k}}{'}_{3}} F({{\bf{k}}'_{3}}) \, \hat{e}_{x_si'_3} T_{i'_3 j'_3}({\bf \hat{k}}'_3) \xi_{j'_3}({\bf{k}}'_3) \right) \right> \\
\end{array}
\end{equation}
Following the rules for spectral representation, the $\varphi_s-$dependent component of the emissivity variance is:
\begin{equation} \label{eq:sxx_OK}
\begin{array}{ll}
Var (\varepsilon_{x})&= \int d{\bf{r}} [C_{xx}({\bf{r}})-C_{xx}({\bf{0}})] \propto \left[(2B_{0\bot}\cos\varphi_s)^2 \int d{{\bf{k}}} F^2({\bf{k}}) e^{i{\bf{k}}{\bf{r}}} \hat{e}_{x_si} \hat{e}_{x_sj} T_{ij}({\bf \hat{k}})\right.\\
   &+ \left. 2\left( \int d{{\bf{k}}} F^2({\bf{k}}) e^{i{\bf{k}}{\bf{r}}} \hat{e}_{x_si} \hat{e}_{x_sj} T_{ij}({\bf \hat{k}}) \right)^2\right]
\end{array}
\end{equation}
It consists of linear and quadratic terms. The linear (1st) term is the most important because it dominates the signature and can be analytically retrieved for different turbulence modes. When observing the turbulence on the 2D plane of sky, the 3D vector ${\bf \hat{k}}\equiv({\rm k_x, k_y, k_z})$ is replaced with the 2D vector ${\bf \hat{K}}\equiv({\rm k_x, k_y, 0})$. The linear term is then:
\begin{equation} \label{eq:sxx_orig}
s_{xx}(\varphi_s) \propto (2\cos\varphi_s)^2 \int_{-\pi}^\pi \, d{\varphi}\, F^2_0({\bf \hat{K}}) \hat{e}_{x_si} \hat{e}_{x_sj} T_{ij}({\bf \hat{K}}) ,
\end{equation}
in which $F^2_0({\bf \hat{K}})$ is the angle dependent factor in the tensor power spectrum $F^2({\bf K})$.

There are two types of spectral tensors for the magnetic axisymmetric turbulence, $T_E$ and $T_F$:
\begin{equation} \label{eq:TE}
T_{E,ij}({\bf \hat{k}}) = \delta_{ij}-\hat{k}_i\hat{k}_j ,
\end{equation}
\begin{equation} \label{eq:TF}
T_{F,ij}({\bf \hat{k}}) = \frac{
({\bf \hat{k}} \cdot {\bf \hat{\lambda}})^2\hat{k}_i\hat{k}_j+\hat{\lambda}_i\hat{\lambda}_j-({\bf \hat{k}} \cdot {\bf \hat{\lambda}})(\hat{k}_i\hat{\lambda}_j+\hat{\lambda}_i\hat{k}_j)
}{1-({\bf \hat{k}} \cdot {\bf \hat{\lambda}})^2} .
\end{equation}
The scalar factor in the tensor power spectrum for Alfv\'en modes is:
\begin{equation} \label{eq:alf_F}
 F^2({\bf{k}}) \propto k^{-11/3} F^2_{0,A}({\bf \hat{k}}) ,
\end{equation}
where
\begin{equation} \label{eq:alf_F0}
 F^2_{0,A}({\bf \hat{k}}) = \exp \left\{ -M_A^{-4/3} \max \left[ 1, (k_{g} L)^{1/3} \right] \frac{|{\bf\hat{k}} \cdot {\bf\hat{\lambda}}|}{\left[ 1-({\bf\hat{k}} \cdot {\bf\hat{\lambda}})^2 \right]^{1/3}} \right\} ,
\end{equation}
in which $L$ is the injection scale of the MHD turbulence. The last expression determines the anisotropy of the spectral energy distribution in the Fourier space. This anisotropy is scale independent,
and it imposed to coincide with the Goldreich \& Sridhar 1995 (henceforth GS95) anisotropy at the scale $1/k_g$ \cite{GS95,YL08}{}.
The scale $1/k_g$ is to be interpreted as the scale of the largest eddies that exhibit the GS95 anisotropy.

\vskip 0.1in
\noindent For {\it MS modes}, the spectral tensor is given by:
\begin{equation} \label{eq:fast_T}
T_{ij}({\bf \hat{k}}) = T_{F,ij}({\bf \hat{k}}) .
\end{equation}
The scalar factor in the tensor spectrum for fast modes is:
\begin{equation} \label{eq:fast_F}
 F^2({\bf{k}}) \propto k^{-7/2} F^2_{0,F}({\bf \hat{k}}) ,
\end{equation}
where
\begin{equation} \label{eq:fast_F0}
 F^2_{0,F}({\bf \hat{k}}) =
  \left\{
    \begin{array}{ll}
      \frac{1}{(1+\frac{\beta}{4}\sin^2\theta_{kB})^2}, & \beta \ll 1 \\
      \frac{1-({\bf \hat{k}}\cdot{\bf \hat{\lambda}})^2}{(1+\frac{1}{\beta}\sin^2\theta_{kB})^2}, & \beta \geq 1 .
    \end{array}
  \right.
\end{equation}
where $\theta_{kB}$ is the angle between mean magnetic field and wave direction.

\vskip 0.1in
\noindent The scalar factor in the tensor spectrum for Slow modes is then:
\begin{equation} \label{eq:slow_F}
 F^2({\bf{k}}) \propto k^{-11/3} F^2_{0,A}({\bf \hat{k}}) \times
  \left\{
    \begin{array}{ll}
      (1-({\bf \hat{k}}\cdot{\bf \hat{\lambda}})^2)(1+\frac{\beta}{2}), & \beta \ll 1 \\
      1+\frac{2}{\beta}, & \beta \geq 1 .
    \end{array}
  \right.
 ,
\end{equation}
where $F^2_{0,A}({\bf{k}})$ is the same as that of Alfv\'en modes (see Eq.\ref{eq:alf_F0}).

Being axisymmetric, the scalar part of the power spectrum $F^2$ for all different modes are even function as shown in Equation~\eqref{eq:alf_F0}~\eqref{eq:fast_F0}. As a result, the non-zero component under the integration $\int_{-\pi}^{\pi}d\varphi$ for the 2D axisymmetric tensors is:
\begin{equation} \label{eq:TExx}
T_{E,xx}({\bf \hat{K}}) = (1-2\sin^2\varphi)\sin^2\varphi_s + \sin^2\varphi,
\end{equation}
\begin{equation} \label{eq:TFxx}
T_{F,xx}({\bf \hat{K}}) = \frac{\sin^2\theta_\lambda\sin^2\varphi(1-2\sin^2\varphi)}{1-\sin^2\theta_\lambda\cos^2\varphi}\sin^2\varphi_s + \frac{\sin^2\theta_\lambda\sin^4\varphi}{1-\sin^2\theta_\lambda\cos^2\varphi}.
\end{equation}
Moreover, considering the local mean field is wandering around the global mean field, $T_{F,ij} $ term is replaced by the global tensor: $T_{F,ij}  \rightarrow W_i T_{E,ij}  + W_l T_{F,ij}$, where $W_i$ and $W_l$ are the weight factors for the isotropic and anisotropic contribution of the turbulence \cite{LP12}{}. Therefore, the signature can be expressed in the format of ``$s_{xx}(\varphi_s)=(a_{xx}\sin^2\varphi_s+b_{xx})\cos^2\varphi_s, \; \varphi_s \in [0,\pi] ,\nonumber$'', where the parameters $a_{xx}$ and $b_{xx}$ for different types of spectral tensors are functions of the tensor spectrum. The $r_{xx}\equiv a_{xx}/b_{xx}$ is identified as ``the classification parameter''. {\it $r_{xx}$ is our target parameter, which is used to
recover the information about the plasma modes composition of turbulence.}
The theoretical analysis of $r_{xx}$ is plotted in Extended Data Fig. 1.

\paragraph*{Observational Analysis}

In this section, we illustrate how to extract the classification parameter of the corresponding turbulence modes from a synchrotron polarization map.
The raw signature $\hat{s}_{xx}$ is first calculated for a spot on the polarization map. The size of each spot is no larger than the coherence length of interstellar magnetic field. An intensity range is selected in order to mask out the background extragalactic point sources. Moreover, high/low-pass gaussian filters are applied in order to remove the scales that are subjected to large-scale fluctuations and instrumental contaminations. The scale of the smallest eddy after the filter should be at least twice the resolution scale (beam size) of the observation. As discussed in the main text, the signature $s_{xx}$, as a function of $\varphi_s$, is composed of linear $\varphi_s$-dependent part and nonlinear part. We first check the percentage of apparent linear signature in total signature (see Figure 1c):
\begin{equation} \label{eq:ksi_xx}
\xi_{xx}=\frac{Amp_{linear}}{Amp_{total}}=\frac{\hat{s}_{xx\max}-\hat{s}_{xx\min}}{\hat{s}_{xx\max}+\hat{s}_{xx\min}}.
\end{equation}
If $\xi_{xx}\leq0.15$, the observed is considered isotropic turbulence. Only the signatures dominated by linear component are accepted, i.e., $\xi_{xx}>0.5$ for classification. Based on the format of linear signature, the ``intrinsic'' linear signature $s_{xx}$ is equal to $0$ at $\varphi_s=90^{\circ}$. The nonlinear part of the raw signature $\hat{s}_{xx}$ is from the quadratic term, the data noise, and free-free emission. The major noise results from the fluctuation of the symmetry axis, which is composed of two parts. One originates from the variation of the local mean magnetic field direction along the line of sight (the line-of-sight component). The other arises from the variation of the local mean magnetic field among different points within the spot (the picture plane component).
We model the influence of the symmetry axis fluctuations on the signature linear term as follows:
\begin{equation} \label{eq:sxx_fluct}
s^{fluct}_{xx}(\varphi_s)= \frac{1}{2 \Delta \varphi_s} \int_{-\Delta \varphi_s}^{\Delta \varphi_s} s_{xx}(\varphi_s-\varphi_s') d{\varphi_s'}
\end{equation}
\noindent where $\Delta \varphi_s$ is the fluctuation angle.
After such operation the signature value at $90^\circ$ is not zero any more:
\begin{equation} \label{eq:sxx_fluct_90}
s^{fluct}_{xx}(90^\circ)= \frac{1}{32\Delta \varphi_s} (4(a_{xx} + 4b_{xx})\Delta \varphi_s - 8b_{xx}\sin 2\Delta \varphi_s - a_{xx}\sin 4\Delta \varphi_s)
\end{equation}
The signature at an arbitrary $\varphi_s$ can be obtained similarly according to Eq. (\ref{eq:sxx_fluct}). The signature in the form of Eq. (\ref{eq:sxx}) is then given by the parameters expressed through original $a_{xx}$ and $b_{xx}$:
\begin{equation} \label{eq:abhat}
\begin{array}{l}
\hat{a}_{xx} = a_{xx} \frac{\sin 4\Delta \varphi_s}{4\Delta \varphi_s} \\
\hat{b}_{xx} = b_{xx} \frac{\sin 2\Delta \varphi_s}{2\Delta \varphi_s}
\end{array}
\end{equation}

Then $\hat{a}_{xx}$ and $\hat{b}_{xx}$ are calculated from the observed signature, use Eqs. (\ref{eq:abhat}) and (\ref{eq:sxx_fluct_90}) to find $\Delta \varphi_s$ and restore the original $r_{xx}$ as follows:
\begin{equation} \label{eq:rxx_rest}
r_{xx} = \hat{r}_{xx} \frac{2 \sin 2\Delta \varphi_s}{\sin 4\Delta \varphi_s}
\end{equation}

The free-free emission is not polarized and therefore has no contribution to the $\varphi_s$ dependent linear signature. The percentage $p$ of fluctuation term in $\hat{s}_{xx}(90^\circ)$ is then scanned from $0$ to $100\%$ (see Figure 1c), yielding a range of $r_{xx}$ values. This value range provides the error budget and the mean value of the classification parameter $r_{xx}$. We reject the spots whose $r_{xx}$ value range crosses the thresholds, i.e. $0,\, 2/3$. We also reject the spots whose $r_{xx}$ is negative and has a value range larger than $0.5$ to constrain the nonlinear component in the signature.

\paragraph*{The Link between rxx regime and plasma modes signature}

Based on the theoretical analysis and the simulation results on decomposed modes, the $r_{xx}$ between -1 and -1/3 is expected in both Alfv\'en and MS modes.
The positive value for $r_{xx}$ (``Alf\'enic'' signature in our recipe) is only expected in trans-Alfv\'enic Alfv\'en modes. On the other hand, the $r_{xx}$ dominantly residing between -1/3 and 0 (``MS'' signature in our recipe) is only expected in MS modes. Additionally, the positive $r_{xx}$ is not expected in the low-$M_A$ regime. In current work, we focus on the ``MS'' and ``Alf\'enic'' signatures, which, based on the numerical simulation, pinpoints the presence of the corresponding modes (see Figure 2). The potential for the $r_{xx}$ behaviour in other ranges will be exploited in future research.

\paragraph*{Axisymmetry}
As demonstrated in the earlier paragraph, the linear signature should be axisymmetric to the mean magnetic field direction at $\varphi_s=90^\circ$ in MHD turbulence. Therefore, the signature is only accepted when satisfying (1) the minimum value of $s_{xx}$ is near $90^\circ$, $\mid{\varphi_s}_{min}-90^\circ\mid\leq5^\circ$; (2) when decompose the signature $s_{xx}(\varphi_s)$ in Fourier space of $\varphi_s$, the $\cos$ term should dominate, i.e.,
\begin{equation}
y_{xx}\equiv\frac{\sqrt{(\int_{-\pi}^{\pi}d{\bf {\varphi_s}}s_{xx}\sin2\varphi_s)^2+(\int_{-\pi}^{\pi}d{\bf{\varphi_s}}s_{xx}\sin4\varphi_s)^2}}{ (s_{xx}(0^\circ)+s_{xx}(90^\circ))/2}<0.2.
\nonumber
\end{equation}

\paragraph*{Synthetic Observations}

In this section, we discuss how synthetic observations are performed with numerical analysis on our MHD turbulence data cubes. Non-thermal electrons with a fixed index are injected into the 3D turbulence data cubes and the resulting synchrotron polarizations are computed along $200$ different line-of-sight directions (LOS). All the turbulence data cubes are $512^3$ with an inertial range more than a decade. In our simulation, two types of Faraday rotation (FR) are accounted for. (1) FR within the emitting layer: The polarization of the synchrotron radiation rotates as it transports along the LOS within the emitting layer. The rotational angle is calculated with respect to the turbulence density and the LOS component of turbulent magnetic field. (2) FR due to a foreground screen: We further propagate the synchrotron radiation from these MHD turbulence data cubes through 10 layers of trans-Alfv\'enic ($M_A\simeq1$) MHD turbulence screen with randomized mean magnetic field orientations. These screens do not produce radiation, but only rotate the polarization of the synchrotron radiation.

Additionally, numerical simulation are also carried out to evaluate the influence of the noise and beam smoothing from observation. First white noise is introduced into our resulting synchrotron I/Q/U maps in simulation. The amplitude of the noise is prescribed according to the Signal-to-Noise ratio (S/Nr) in the environments under study. In Urumqi survey, the actual noise amplitude from observation is $0.9mK$ for Stokes-I and $0.6mK$ for Stokes-Q/U \cite{Xiao2011}{}. The S/Nr in Stokes-I map is calculated based on the ratio between the average intensity of the region and the noise amplitude, i.e., $S/Nr^{I}=I_{rms}/\Delta{I}\sim 8$. For Stokes-Q/U, The S/Nr is computed for total polarization ($P\equiv \sqrt{Q^2+U^2}$) and then divide it by $\sqrt{2}$ to get the S/Nr for Q or U. Following that, the beam smoothing is performed with real observational scales. In our observational analysis, the ratio between largest scale eddy and the beam size is $\sim4$. The beam size in the numerical simulation is determined accordinly. Taking the beam size as Full width half maximum (FWHM) of the signal, the synchrotron I/Q/U maps is smoothed with a 2D-Gaussian beam, whose sigma scale is determined by $l_{beam}=\sqrt{8 ln2}\sigma$. As demonstrated in Figure 2f, the dominant signature reveals the corresponding plasma modes unambiguously.

\paragraph*{Data availability.} We used the synchrotron polarization data from Urumqi 6 cm polarization survey\\
(\url{https://www3.mpifr-bonn.mpg.de/survey.html}).

\paragraph*{Code availability.} We opt not to make the SPA code publicly available now because it is still under development and maintenance. We intend to publish the code at a later time. The codes used to generate the plots presented in this paper are available from the corresponding author upon reasonable request.


\begin{thebibliography}{10}
\expandafter\ifx\csname url\endcsname\relax
  \def\url#1{\texttt{#1}}\fi
\expandafter\ifx\csname urlprefix\endcsname\relax\def\urlprefix{URL }\fi
\providecommand{\bibinfo}[2]{#2}
\providecommand{\eprint}[2][]{\url{#2}}

\bibitem{Armstrong95}
\bibinfo{author}{{Armstrong}, J.~W.}, \bibinfo{author}{{Rickett}, B.~J.} \&
  \bibinfo{author}{{Spangler}, S.~R.}
\newblock \bibinfo{title}{{Electron Density Power Spectrum in the Local
  Interstellar Medium}}.
\newblock \emph{\bibinfo{journal}{Astrophys. J.}}
  \textbf{\bibinfo{volume}{443}}, \bibinfo{pages}{209} (\bibinfo{year}{1995}).

\bibitem{Elmegreen2004}
\bibinfo{author}{{Elmegreen}, B.~G.} \& \bibinfo{author}{{Scalo}, J.}
\newblock \bibinfo{title}{{Interstellar Turbulence I: Observations and
  Processes}}.
\newblock \emph{\bibinfo{journal}{Annu. Rev. Astron. Astrophys.}}
  \textbf{\bibinfo{volume}{42}}, \bibinfo{pages}{211--273}
  (\bibinfo{year}{2004}).

\bibitem{GS95}
\bibinfo{author}{{Goldreich}, P.} \& \bibinfo{author}{{Sridhar}, S.}
\newblock \bibinfo{title}{{Toward a theory of interstellar turbulence. 2:
  Strong alfvenic turbulence}}.
\newblock \emph{\bibinfo{journal}{Astrophys. J.}}
  \textbf{\bibinfo{volume}{438}}, \bibinfo{pages}{763--775}
  (\bibinfo{year}{1995}).

\bibitem{LG01}
\bibinfo{author}{{Lithwick}, Y.} \& \bibinfo{author}{{Goldreich}, P.}
\newblock \bibinfo{title}{{Compressible Magnetohydrodynamic Turbulence in
  Interstellar Plasmas}}.
\newblock \emph{\bibinfo{journal}{Astrophys. J.}}
  \textbf{\bibinfo{volume}{562}}, \bibinfo{pages}{279--296}
  (\bibinfo{year}{2001}).

\bibitem{CL03}
\bibinfo{author}{{Cho}, J.} \& \bibinfo{author}{{Lazarian}, A.}
\newblock \bibinfo{title}{{Compressible magnetohydrodynamic turbulence: mode
  coupling, scaling relations, anisotropy, viscosity-damped regime and
  astrophysical implications}}.
\newblock \emph{\bibinfo{journal}{Mon. Not. R. Astron. Soc.}}
  \textbf{\bibinfo{volume}{345}}, \bibinfo{pages}{325--339}
  (\bibinfo{year}{2003}).

\bibitem{MY19}
\bibinfo{author}{{Makwana}, K.~D.} \& \bibinfo{author}{{Yan}, H.}
\newblock \bibinfo{title}{{Properties of magnetohydrodynamic modes in
  compressively driven plasma turbulence}}.
\newblock \emph{\bibinfo{journal}{arXiv e-prints}}
  \bibinfo{pages}{arXiv:1907.01853} (\bibinfo{year}{2019}).

\bibitem{YL02}
\bibinfo{author}{{Yan}, H.} \& \bibinfo{author}{{Lazarian}, A.}
\newblock \bibinfo{title}{{Scattering of Cosmic Rays by Magnetohydrodynamic
  Interstellar Turbulence}}.
\newblock \emph{\bibinfo{journal}{Phys. Rev. Lett.}}
  \textbf{\bibinfo{volume}{89}}, \bibinfo{pages}{281102}
  (\bibinfo{year}{2002}).

\bibitem{Lynn14}
\bibinfo{author}{{Lynn}, J.~W.}, \bibinfo{author}{{Quataert}, E.},
  \bibinfo{author}{{Chandran}, B.~D.~G.} \& \bibinfo{author}{{Parrish}, I.~J.}
\newblock \bibinfo{title}{{Acceleration of Relativistic Electrons by
  Magnetohydrodynamic Turbulence: Implications for Non-thermal Emission from
  Black Hole Accretion Disks}}.
\newblock \emph{\bibinfo{journal}{Astrophys. J.}}
  \textbf{\bibinfo{volume}{791}}, \bibinfo{pages}{71} (\bibinfo{year}{2014}).

\bibitem{MO77}
\bibinfo{author}{{McKee}, C.~F.} \& \bibinfo{author}{{Ostriker}, J.~P.}
\newblock \bibinfo{title}{{A theory of the interstellar medium - Three
  components regulated by supernova explosions in an inhomogeneous substrate}}.
\newblock \emph{\bibinfo{journal}{Astrophys. J.}}
  \textbf{\bibinfo{volume}{218}}, \bibinfo{pages}{148--169}
  (\bibinfo{year}{1977}).

\bibitem{KH10_accretion}
\bibinfo{author}{{Klessen}, R.~S.} \& \bibinfo{author}{{Hennebelle}, P.}
\newblock \bibinfo{title}{{Accretion-driven turbulence as universal process:
  galaxies, molecular clouds, and protostellar disks}}.
\newblock \emph{\bibinfo{journal}{Astron. Astrophys.}}
  \textbf{\bibinfo{volume}{520}}, \bibinfo{pages}{A17} (\bibinfo{year}{2010}).

\bibitem{Balbus99}
\bibinfo{author}{{Sellwood}, J.~A.} \& \bibinfo{author}{{Balbus}, S.~A.}
\newblock \bibinfo{title}{{Differential Rotation and Turbulence in Extended H I
  Disks}}.
\newblock \emph{\bibinfo{journal}{Astrophys. J.}}
  \textbf{\bibinfo{volume}{511}}, \bibinfo{pages}{660--665}
  (\bibinfo{year}{1999}).

\bibitem{KN02_thermal}
\bibinfo{author}{{Kritsuk}, A.~G.} \& \bibinfo{author}{{Norman}, M.~L.}
\newblock \bibinfo{title}{{Thermal Instability-induced Interstellar
  Turbulence}}.
\newblock \emph{\bibinfo{journal}{Astrophys. J.l}}
  \textbf{\bibinfo{volume}{569}}, \bibinfo{pages}{L127--L131}
  (\bibinfo{year}{2002}).

\bibitem{NL07_outflow}
\bibinfo{author}{{Nakamura}, F.} \& \bibinfo{author}{{Li}, Z.-Y.}
\newblock \bibinfo{title}{{Protostellar Turbulence Driven by Collimated
  Outflows}}.
\newblock \emph{\bibinfo{journal}{Astrophys. J.}}
  \textbf{\bibinfo{volume}{662}}, \bibinfo{pages}{395--412}
  (\bibinfo{year}{2007}).

\bibitem{YL08}
\bibinfo{author}{{Yan}, H.} \& \bibinfo{author}{{Lazarian}, A.}
\newblock \bibinfo{title}{{Cosmic-Ray Propagation: Nonlinear Diffusion Parallel
  and Perpendicular to Mean Magnetic Field}}.
\newblock \emph{\bibinfo{journal}{Astrophys. J.}}
  \textbf{\bibinfo{volume}{673}}, \bibinfo{pages}{942--953}
  (\bibinfo{year}{2008}).

\bibitem{slowmodes_shi15}
\bibinfo{author}{{SHI}, M.~J.} \emph{et~al.}
\newblock \bibinfo{title}{{Observations of Alfv{\'e}n and Slow Waves in the
  Solar Wind near 1 AU}}.
\newblock \emph{\bibinfo{journal}{Astrophys. J.}}
  \textbf{\bibinfo{volume}{815}}, \bibinfo{pages}{122} (\bibinfo{year}{2015}).

\bibitem{YL04}
\bibinfo{author}{{Yan}, H.} \& \bibinfo{author}{{Lazarian}, A.}
\newblock \bibinfo{title}{{Cosmic-Ray Scattering and Streaming in Compressible
  Magnetohydrodynamic Turbulence}}.
\newblock \emph{\bibinfo{journal}{Astrophys. J.}}
  \textbf{\bibinfo{volume}{614}}, \bibinfo{pages}{757--769}
  (\bibinfo{year}{2004}).

\bibitem{LP12}
\bibinfo{author}{{Lazarian}, A.} \& \bibinfo{author}{{Pogosyan}, D.}
\newblock \bibinfo{title}{{Statistical Description of Synchrotron Intensity
  Fluctuations: Studies of Astrophysical Magnetic Turbulence}}.
\newblock \emph{\bibinfo{journal}{Astrophys. J.}}
  \textbf{\bibinfo{volume}{747}}, \bibinfo{pages}{5} (\bibinfo{year}{2012}).

\bibitem{PENCILCITE}
\bibinfo{author}{{Brandenburg}, A.} \& \bibinfo{author}{{Dobler}, W.}
\newblock \bibinfo{title}{{Pencil: Finite-difference Code for Compressible
  Hydrodynamic Flows}} (\bibinfo{year}{2010}).

\bibitem{PLUTOCITE}
\bibinfo{author}{{Mignone}, A.} \emph{et~al.}
\newblock \bibinfo{title}{{PLUTO: A Numerical Code for Computational
  Astrophysics}}.
\newblock \emph{\bibinfo{journal}{Astrophys. J. S.}}
  \textbf{\bibinfo{volume}{170}}, \bibinfo{pages}{228--242}
  (\bibinfo{year}{2007}).

\bibitem{Xiao2011}
\bibinfo{author}{{Xiao}, L.} \emph{et~al.}
\newblock \bibinfo{title}{{A Sino-German {\ensuremath{\lambda}}6 cm
  polarization survey of the Galactic plane. IV. The region from 60 to 129
  degree longitude}}.
\newblock \emph{\bibinfo{journal}{Astron. Astrophys.}}
  \textbf{\bibinfo{volume}{529}}, \bibinfo{pages}{A15} (\bibinfo{year}{2011}).

\bibitem{Opper2015}
\bibinfo{author}{{Oppermann}, N.} \emph{et~al.}
\newblock \bibinfo{title}{{Estimating extragalactic Faraday rotation}}.
\newblock \emph{\bibinfo{journal}{Astron. Astrophys.}}
  \textbf{\bibinfo{volume}{575}}, \bibinfo{pages}{A118} (\bibinfo{year}{2015}).

\bibitem{Schneider2011}
\bibinfo{author}{{Schneider}, N.} \emph{et~al.}
\newblock \bibinfo{title}{{The link between molecular cloud structure and
  turbulence}}.
\newblock \emph{\bibinfo{journal}{Astron. Astrophys.}}
  \textbf{\bibinfo{volume}{529}}, \bibinfo{pages}{A1} (\bibinfo{year}{2011}).

\bibitem{COCompare}
\bibinfo{author}{{Dame}, T.~M.}, \bibinfo{author}{{Hartmann}, D.} \&
  \bibinfo{author}{{Thaddeus}, P.}
\newblock \bibinfo{title}{{The Milky Way in Molecular Clouds: A New Complete CO
  Survey}}.
\newblock \emph{\bibinfo{journal}{Astrophys. J.}}
  \textbf{\bibinfo{volume}{547}}, \bibinfo{pages}{792--813}
  (\bibinfo{year}{2001}).

\bibitem{FermiLAT:2011}
\bibinfo{author}{{Ackermann}, M.} \emph{et~al.}
\newblock \bibinfo{title}{{A Cocoon of Freshly Accelerated Cosmic Rays Detected
  by Fermi in the Cygnus Superbubble}}.
\newblock \emph{\bibinfo{journal}{Science}} \textbf{\bibinfo{volume}{334}},
  \bibinfo{pages}{1103} (\bibinfo{year}{2011}).

\bibitem{Beerer10}
\bibinfo{author}{{Beerer}, I.~M.} \emph{et~al.}
\newblock \bibinfo{title}{{A Spitzer View of Star Formation in the Cygnus X
  North Complex}}.
\newblock \emph{\bibinfo{journal}{Astrophys. J.}}
  \textbf{\bibinfo{volume}{720}}, \bibinfo{pages}{679--693}
  (\bibinfo{year}{2010}).

\bibitem{Herschell16}
\bibinfo{author}{{Schneider}, N.} \emph{et~al.}
\newblock \bibinfo{title}{{Globules and pillars in Cygnus X. I. Herschel
  far-infrared imaging of the Cygnus OB2 environment}}.
\newblock \emph{\bibinfo{journal}{Astron. Astrophys.}}
  \textbf{\bibinfo{volume}{591}}, \bibinfo{pages}{A40} (\bibinfo{year}{2016}).

\bibitem{Wright12}
\bibinfo{author}{{Wright}, N.~J.} \emph{et~al.}
\newblock \bibinfo{title}{{Photoevaporating Proplyd-like Objects in Cygnus
  OB2}}.
\newblock \emph{\bibinfo{journal}{Astrophys. J. Lett.}}
  \textbf{\bibinfo{volume}{746}}, \bibinfo{pages}{L21} (\bibinfo{year}{2012}).

\bibitem{Rygl12}
\bibinfo{author}{{Rygl}, K.~L.~J.} \emph{et~al.}
\newblock \bibinfo{title}{{Parallaxes and proper motions of interstellar masers
  toward the Cygnus X star-forming complex. I. Membership of the Cygnus X
  region}}.
\newblock \emph{\bibinfo{journal}{Astron. Astrophys.}}
  \textbf{\bibinfo{volume}{539}}, \bibinfo{pages}{A79} (\bibinfo{year}{2012}).

\bibitem{Maia16}
\bibinfo{author}{{Maia}, F.~F.~S.}, \bibinfo{author}{{Moraux}, E.} \&
  \bibinfo{author}{{Joncour}, I.}
\newblock \bibinfo{title}{{Young and embedded clusters in Cygnus-X: evidence
  for building up the initial mass function?}}
\newblock \emph{\bibinfo{journal}{Mon. Not. R. Astron. Soc.}}
  \textbf{\bibinfo{volume}{458}}, \bibinfo{pages}{3027--3046}
  (\bibinfo{year}{2016}).

\bibitem{HAWCCYG}
\bibinfo{author}{{Abeysekara}, A.~U.} \emph{et~al.}
\newblock \bibinfo{title}{{The 2HWC HAWC Observatory Gamma-Ray Catalog}}.
\newblock \emph{\bibinfo{journal}{Astrophys. J.}}
  \textbf{\bibinfo{volume}{843}}, \bibinfo{pages}{40} (\bibinfo{year}{2017}).

\bibitem{Odegard1986}
\bibinfo{author}{{Odegard}, N.}
\newblock \bibinfo{title}{{Decameter Wavelength Observations of the Rosette
  Nebula and the Monoceros Loop Supernova Remnant}}.
\newblock \emph{\bibinfo{journal}{Astrophys. J.}}
  \textbf{\bibinfo{volume}{301}}, \bibinfo{pages}{813} (\bibinfo{year}{1986}).

\end{thebibliography}

\begin{thebibliography}{10}
\expandafter\ifx\csname url\endcsname\relax
  \def\url#1{\texttt{#1}}\fi
\expandafter\ifx\csname urlprefix\endcsname\relax\def\urlprefix{URL }\fi
\providecommand{\bibinfo}[2]{#2}
\providecommand{\eprint}[2][]{\url{#2}}
\setcounter{enumiv}{31}

\bibitem{R90}
\bibinfo{author}{{Rozanov}, Y.~A.}
\newblock In \emph{\bibinfo{booktitle}{Stationary Random Processes}}
  (\bibinfo{publisher}{Nauka, Moscow}, \bibinfo{year}{1990}).

\bibitem{Che98}
\bibinfo{author}{{Chepurnov}, A.~V.}
\newblock \bibinfo{title}{{The Galactic Foreground Angular Spectra}}.
\newblock \emph{\bibinfo{journal}{Astron. Astrophys. Trans.}}
  \textbf{\bibinfo{volume}{17}}, \bibinfo{pages}{281--300}
  (\bibinfo{year}{1998}).

\end{thebibliography}
\end{document}